\documentclass[grl]{agu2001}
\usepackage{agu-ps}
\usepackage[dvips,final]{graphicx}
\ifgalley
\setlastpagenum{4}
\fi
\newcommand{\be}{\begin{equation}}
\newcommand{\ee}{\end{equation}}

\newcommand{\paref}[1]{(\ref{#1})}
\newcommand{\eqref}[1]{eq.\,(\ref{#1})}
\newcommand{\Eqref}[1]{Eq.\,(\ref{#1})}
\newcommand{\abs}[1]{\left|#1\right|}
\newenvironment{eq}{\vspace{2mm}\begin{equation}\begin{array}{l}}{\end{array}
                    \end{equation}\vspace{2mm}}
\newcommand{\figwidth}{\hsize}
\newcommand{\figref}[1]{Fig.\,\ref{#1}}
\newcommand{\tabref}[1]{Tab.\,\ref{#1}}
\authorrunninghead{COSTA AND MACEDONIO }
\titlerunninghead{HYSTERESIS EFFECTS IN MAGMA FLOWS}

\journalid{}
\articleid{}{}
\paperid{2001GL014493}
\cpright{agu}{2002}
\received{November 30, 2001}
\revised{}
\accepted{}
\published{}

\authoraddr{A. Costa,
Dipartimento di Scienze della Terra e Geologico-Ambientali,
Universit\`a di Bologna, Via Zamboni 67, I-40126 Bologna, Italy.
(e-mail: costa@ov.ingv.it)}

\authoraddr{G. Macedonio,
Osservatorio Vesuviano - INGV,
Via Diocleziano 328, I-80124 Napoli, Italy.
(e-mail: \hbox{macedon@}ov.ingv.it)}

\begin{document}
\title{Nonlinear phenomena in fluids with temperature-dependent
viscosity: an hysteresis model for magma flow in conduits}

\author{Antonio Costa\altaffilmark{1}}
\affil{Dipartimento di Scienze della Terra e Geologico-Ambientali,
Universit\`a di Bologna, Italy}

\altaffiltext{1}{Also at Dipartimento di Scienze della Terra, Universit\`a
degli Studi di Pisa, Italy.}

\author{Giovanni Macedonio}
\affil{Osservatorio Vesuviano, Istituto Nazionale di Geofisica e
Vulcanologia, Napoli, Italy}
\begin{abstract}
Magma viscosity is strongly temperature-dependent. When hot magma flows in
a conduit, heat is lost through the walls and the temperature
decreases along the flow causing a viscosity increase.
For particular
values of the controlling parameters the steady-flow regime in a
conduit shows two
stable solutions belonging either to the slow or to the fast branch. As a
consequence, this
system may show an hysteresis effect, and the transition between the
two branches can occur quickly when certain critical points are reached. 
In this paper we describe a model to
study the relation between the pressure at the inlet and the
volumetric magma flow rate in a conduit.
We apply this model to explain an hysteric jump observed during the
dome growth at Soufri\`ere Hills volcano (Montserrat), and
described by Melnik and Sparks [1999] using a different model.
\end{abstract}
\begin{article}
\section{Introduction}
Like other liquids of practical interest (e.g.\ polymers,
oils), magma has a strong temperature-dependent viscosity.
It is well known that the flow in conduits of this kind of fluids
admits one or more solutions for different ranges of the controlling
parameters.
The  problem of the multiple solutions and their stability was studied by
\cite{peasha73} and by \cite{alenai79}. 
More recently this phenomenon, applied to magma flows, was
investigated with more sophisticated models by \cite{hel95} and by
\cite{wyllis95}.
In \cite{skusla99} a simple method to find the conditions of multiple
solutions was adopted and an experimental study to verify the hysteresis 
phenomenon predicted for this process was performed.
\section{The model}
In this paper we investigate a simple one-dimensional flow
model of a fluid with temperature-dependent viscosity, with the
essential physical properties characterizing the
phenomenon of the multiple solutions and hysteresis as in
\cite{skusla99}.
The fluid flows in a conduit with constant cross section and constant
temperature at the wall boundaries.
We assume that the fluid properties are constant with the temperature
except for the viscosity, and we neglect the heat conduction along the
streamlines, and the viscous heat generation. Moreover, we assume a
linear relation between the shear stress and the strain rate
(Newtonian rheology). This last assumption is introduced to simplify
the model and allows us to demonstrate that the multiple solutions in
conduit flows are the direct consequence of the viscosity increase
along the conduit induced by cooling (under particular boundary conditions).
Under these hypotheses, the equations for momentum and energy balance for the
one-dimensional steady flow in a long circular
conduit ($R \ll L$) at low Reynolds number are:
\be
\frac{1}{r}\frac{\partial }{\partial r}\left(\mu(T) r\frac{\partial
v}{\partial r}\right)=\frac{d P}{d z}
\label{eq:stress}
\ee
\be
\frac{1}{r}\frac{\partial }{\partial r}\left(k r\frac{\partial
T}{\partial r}\right)=\rho c_p v \frac{\partial T}{\partial z} 
\label{eq:energia}
\ee
where $R$ is the conduit radius, $L$ conduit length, $r$ 
radial direction, $z$ direction along the flow, $v$
velocity along the flow (we assume $v=v(r,z)\approx v(r)$), $c_p$
specific heat at constant pressure, $k$ thermal conductivity,
$\rho$ density, $\mu$ viscosity, and $P$ pressure, and $T$ temperature.
For magma, the dependence of the viscosity on the temperature is
well described by the Arrhenius law:
\be
\mu=\mu_A \exp(B/T)
\label{eq:arrh}
\ee
where $\mu_A$ is a constant and $B$ the activation energy. 
In this paper, for simplicity, we approximate \eqref{eq:arrh} by the
Nahme's exponential law, valid when $(T-T_R)/T_R \ll 1$, where $T_R$
is a reference temperature:
\be
\mu=\mu_R \exp\left[-\beta(T-T_R)\right]
\label{eq:exp}
\ee
with $\beta=B/T_R^2$ and $\mu_R=\mu_A\exp(B/T_R)$.
Following \cite{skusla99},
we introduce two new variables: the volumetric flow rate
\be
Q=2\pi \int_0^R v(r) r dr
\label{eq:flux} 
\ee
and the convected mean temperature:
\be
T^*(z)=\frac{2\pi}{Q}\int_0^R T(r,z)v(r) r dr
\label{eq:tmedia}
\ee
To satisfy the mass conservation, the volumetric flow rate $Q$ is
constant along the flow.
Integrating \eqref{eq:stress} and \paref{eq:energia} and
expressing the solutions in terms of $Q$ and $T^*$, we obtain:
\be
\frac{\pi R^4 \exp{[\beta(T^*-T_w)]}}{8\mu_w}\abs{\frac{dP}{dz}}=Q
\label{eq:q1}
\ee
\be
\rho c_p Q\frac{\partial T^*}{\partial z}=2\pi R\alpha(T_w-T^*)
\label{eq:t1}
\ee
where $T_w$ is the wall temperature (taken here as reference,
$T_w=T_R$), $\mu_w$ fluid viscosity at $T=T_w$, and $\alpha$
the coefficient of heat transfer through the walls defined as:
\be
\alpha=\left.\frac{k}{T_w-T^*}\ \frac{\partial T}{\partial r}\right|_{r=R}
\label{eq:alfa}
\ee
In the averaged model (eq.\,\ref{eq:q1}), we have adopted
\be
\mu\approx \mu_R \exp\left[-\beta(T^*-T_R)\right]
\label{eq:exp2}
\ee
and as pointed out in \cite{hel95}, the value of $\beta$ in  
\eqref{eq:exp2}, is usually smaller than the actual value of
$\beta$ in the fluid (eq.\,\ref{eq:exp}).
\par
\begin{figure}
\noindent\includegraphics[angle=-90,width=\figwidth]{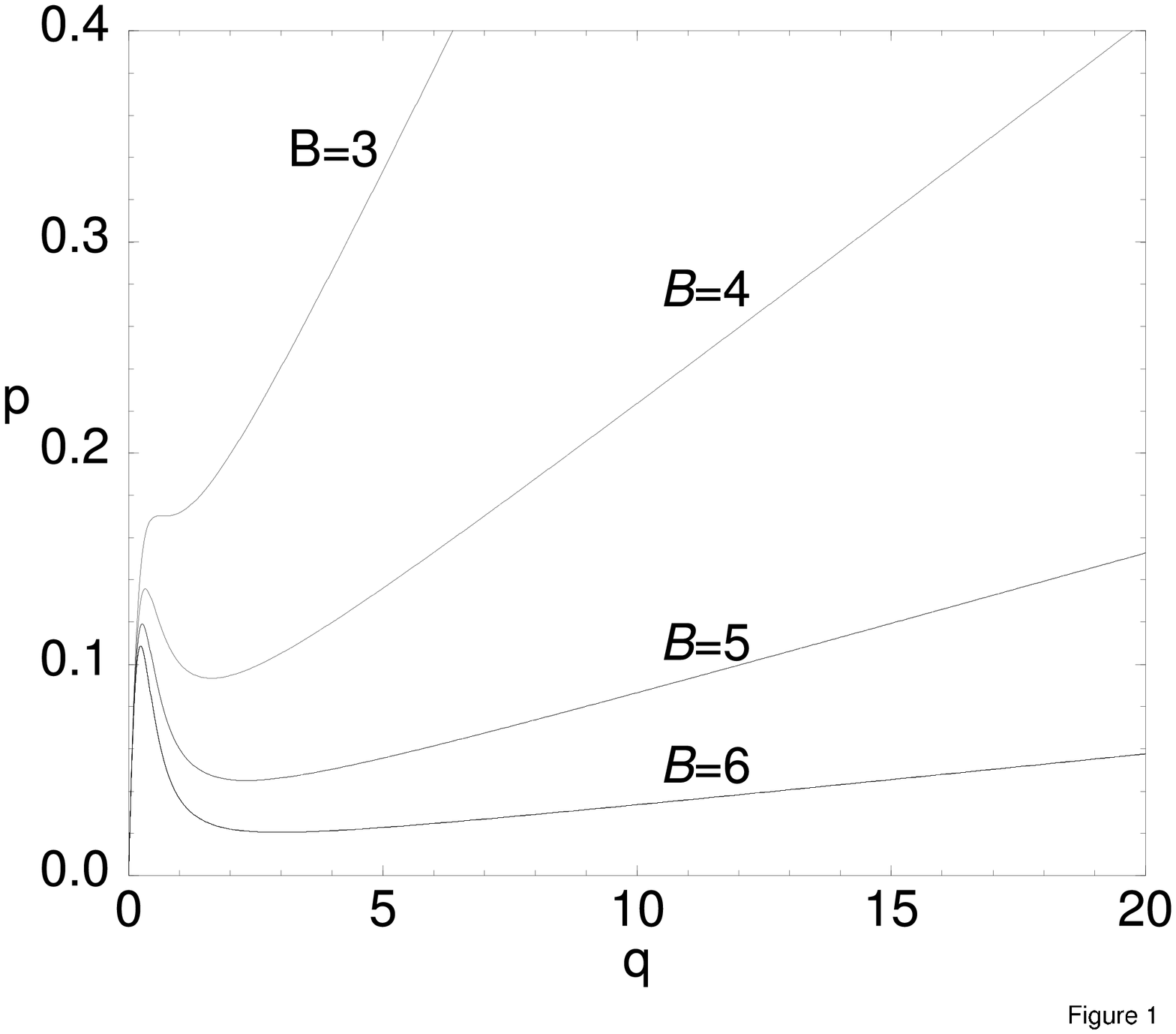}
\caption{Plot of the relation between the nondimensional flow rate $q$
and the nondimensional pressure at conduit inlet $p$, for different
values of the parameter ${\cal B}$, resulting from \eqref{eq:solution}.}
\label{fig:pq}
\end{figure}

\Eqref{eq:q1} and \paref{eq:t1} with the boundary conditions:
\be
P(0)=P_0, \quad P(L)=0, \quad T^*(0)=T_0
\label{eq:bcdim}
\ee
give an approximate solution of the problem. These equations are
similar to the model of \cite{peasha73} (for a plane
flow) and were used by \cite{shapea74} to study the
viscous heating effects. In the nondimensional form, \eqref{eq:q1} and
\paref{eq:t1} are:
\be
\begin{array}{ll}
\displaystyle
\frac{dp}{d\zeta}= - q e^{-\theta} \\[3ex]
\displaystyle
q\frac{d\theta}{d\zeta}+\theta=0
\label{eq:q2}
\end{array}
\ee
where
\be
\begin{array}{lcl}
\displaystyle
q=\frac{\rho c_pQ}{2\pi RL\alpha},& \qquad &
\zeta=\displaystyle\frac{z}{L}\\[2ex]
\displaystyle
p=\frac{\rho c_p R^3}{16\mu_w L^2\alpha}P, & \qquad &
\theta=\beta(T^*-T_w) \\
\end{array}
\label{eq:adim}
\ee
The boundary conditions \paref{eq:bcdim} are rewritten for the new
variables:
\be
p(0)=p_0=\frac{\rho c_p R^3 P_0}{16\mu_w L^2\alpha}, \quad p(1)=0,
\quad \theta(0) = {\cal B}
\label{eq:bcadim}
\ee
with ${\cal B}=\beta(T_0-T_w)$.
The solutions of \eqref{eq:q2}, satisfying
\eqref{eq:bcadim} at the boundaries, are:
\begin{eq}
p(\zeta)-p_0=q\int^0_\zeta\exp(-{\cal B}e^{-\zeta/q})d\zeta \\[2ex]
\theta(\zeta)={\cal B}\exp(-\zeta/q)
\label{eq:sols}
\end{eq}
and, therefore, the relation between the nondimensional pressure at the
conduit inlet $p_0$ and the nondimensional flow rate $q$ is: 
\be
p_0=q\int^1_0\exp(-{\cal B}e^{-\zeta/q})d\zeta
\label{eq:solution}
\ee
In \callout{\figref{fig:pq}} we plot relation \paref{eq:solution} obtained
numerically. We observe that for values of ${\cal B}$ greater
than a critical value ${\cal B}_c\simeq 3$, there are values of $p_0$
which correspond to three different values of $q$.  
By using a simpler model, \cite{skusla99} found ${\cal B}_c=4$, whereas 
\cite{hel95} found ${\cal B}_c=3.03$, in good agreement with \cite{peasha73}.
Moreover, \cite{shapea74} showed that when the viscous heat generation
is important, the value of ${\cal B}_c$ can be lower, but for high
values of ${\cal B}$, the relation between $p_0$ and $q$ is similar
to the case without viscous heat generation.
\par
The stability analysis of the three branches (slow,
intermediate and fast) shows that the intermediate branch is never
stable. Moreover, one part of the slow branch is stable to
two-dimensional perturbations but unstable to the three-dimensional ones,
in a way similar to the Saffman-Taylor instability
\citep{wyllis95}.
In the case of multiple solutions, an hysteresis
phenomenon may occur, as proposed in \cite{wyllis95} and verified
experimentally in \cite{skusla99}.
\par
In the experiments of \cite{skusla99} a fluid with prescribed
temperature and pressure was injected into a capillary tube with
constant wall temperature and controlled fluid pressure and flow rate.
The device is used to show the
transition between the two regimes corresponding to the upper and the
lower branches.
\par
\begin{figure}
\noindent\includegraphics[angle=-90,width=\figwidth]{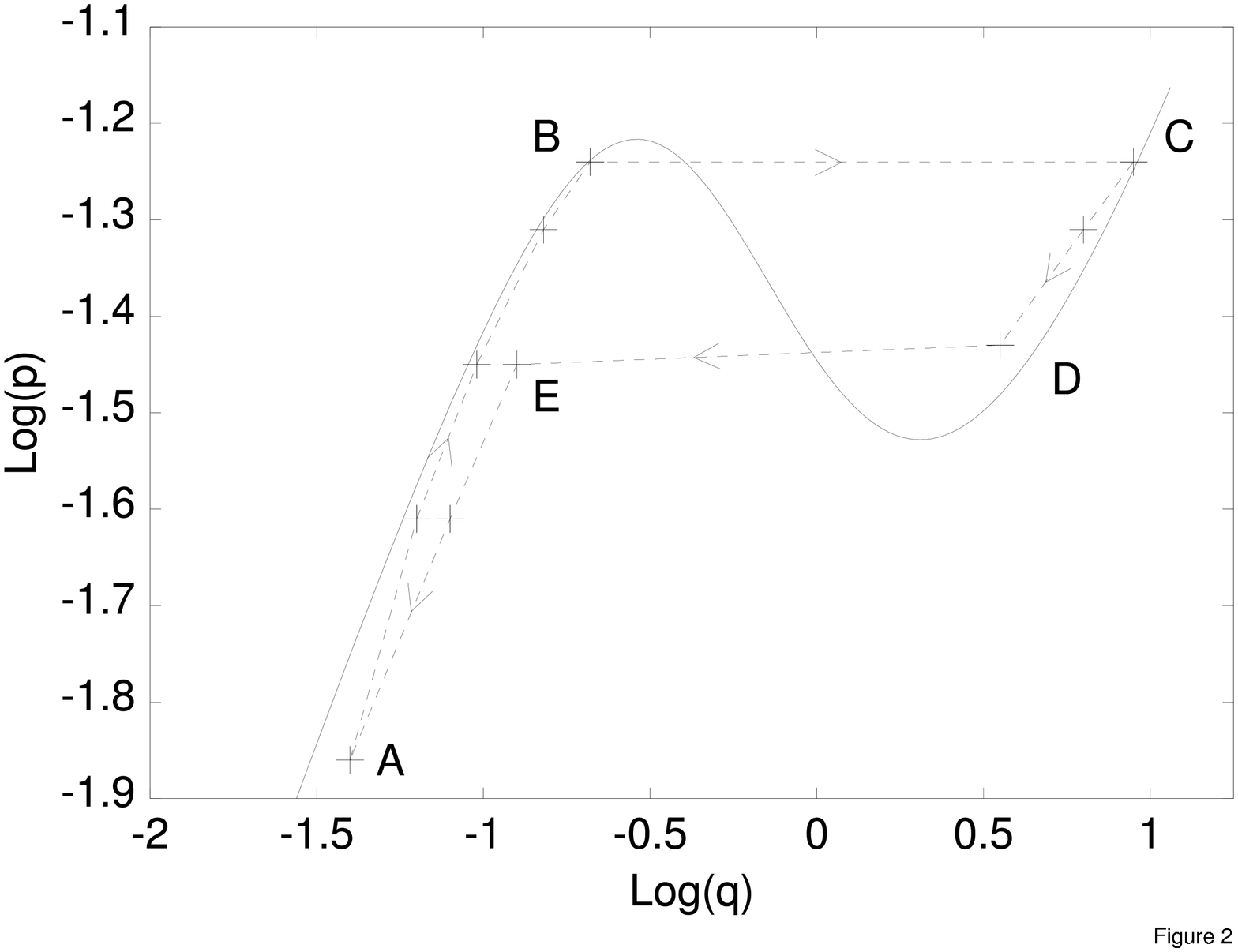}
\caption{Plot of the experimental results (crosses) of \cite{skusla99}.
The dashed
line follows the ``history'' of the flow regimes imposed to the fluid;
the full line represents the path predicted by the model with the same
parameters of the experiment. Modified after \cite{skusla99}.}
\label{fig:esp}
\end{figure}

A comparison between the experimental results (crosses) and theory
(full line) is shown in \callout{\figref{fig:esp}} for the
nondimensional variables $p_0$ and $q$, for ${\cal B}=4.6$.
The dashed lines indicate the
pressure history prescribed in the experiments. The two
steady-state regimes corresponding to the slow and to the fast branch were
clearly recorded. Starting with a  low pressure configuration 
(point A) and by increasing the pressure, the flow rate increases along
the slow branch until it reaches a critical point (point B). Here,
a jump to the fast branch occurs (point C). Increasing the
pressure further, the flow rate increases along this branch,
whereas, by decreasing the pressure, the flow rate decreases moving along 
the upper branch, until it reaches another critical point (point D) where 
the jump on the slow branch occurs (point E). On the slow
branch the nondimensional flow rate is more than one order of magnitude lower
than that on the fast branch.
\section{Application}
During some basaltic fissure eruptions in Hawaii and in Iceland, the
eruption begins with a rapid opening at high flow rate
and, after few hours, the flow rate quickly decreases. To explain this
phenomenon, \cite{wyllis95} and \cite{hel95} proposed a model similar
to the model presented in this paper, based on
the hysteric jump between the fast branch and the slow
branch when the pressure driving the eruption decreases.
\par
A similar phenomenon, showing a jump in the mass flow rate,
was observed during the dome growth (1995-1999) at Soufri\`ere Hills
volcano in Montserrat as described by \cite{melspa99}.
The model used by \cite{melspa99} to explain this effect is essentially based
on the crystal growth kinetics which affects magma viscosity,
and the mechanical coupling between the gas and the melt through the
Darcy law.
\par
In this paper we explain the same phenomenon in
terms of the viscosity variation governed by cooling along the flow.
However, since the crystal content is physically related to the magma
temperature, the two models are physically related.
\par
Using the data of \callout{\tabref{tab:dati}}, we fit the curve of
\eqref{eq:solution} with the observed values of the discharge rates
and dome height reported in \cite{melspa99}. 
\begin{table}
\caption{Typical Values of the Parameters Used in This Paper}
\begin{tabular}{lccl}
\tableline
Parameter & Symbol & Value & Unit\\
\tableline
Rock density &$\rho_r$   & 2600 & Kg m$^{-3}$ \\
Magma density &$\rho_m$  & 2300 & Kg m$^{-3}$ \\
Conduit radius &$R$      & 15   & m \\
Conduit length &$L$ & 5000 & m \\
Magma temperature &$T_0$ & 1123 & K \\
Magma specific heat & $c_p$ & 1000 & J Kg$^{-1}$ K$^{-1}$\\
Magma thermal conductivity& $k$ & 2 & W m$^{-1}$ K$^{-1}$\\
\tableline
\end{tabular}
\label{tab:dati}
\end{table}
The variation of the dome height reflects a change of the exit
pressure and, as a consequence, a variation of the difference between
the inlet and the outlet pressures. Since we assume a zero
exit pressure in our model and the gravity term was not explicitly accounted
for in \eqref{eq:stress},
the variable $p_0$ represents the overpressure at the base of the conduit
(to obtain the actual value, the hydrostatic pressure $\rho g (L+H)$
must be added to its value, where $L$ is the conduit length and $H$ is
the dome height). 
\par
\begin{figure}
\noindent\includegraphics[angle=-90,width=\figwidth]{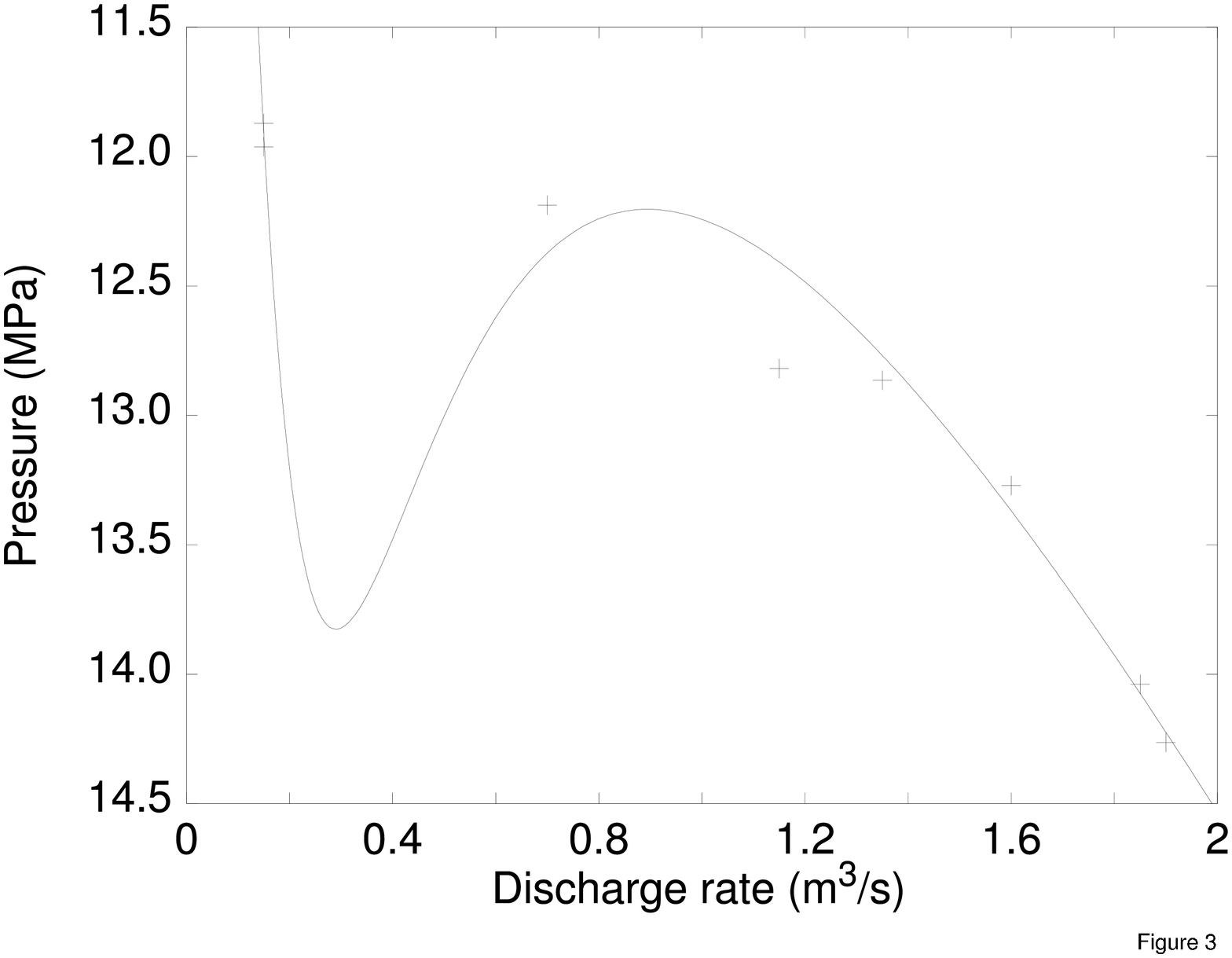}
\caption{Relation between the discharge rate and the pressure. Full
line represents the model prediction; crosses represents
observed values at Soufri\`ere Hills from \cite{melspa99}.}
\label{fig:fitdati}
\end{figure}

The values
of $\alpha$, ${\cal B}$ and $\mu_w$ are chosen by least square best
fitting of the observed data, and the wall temperature $T_w$ was
defined as the temperature for which magma ceases to flow.
\callout{\figref{fig:fitdati}} shows the results of least square fitting:
the discharge rate is reported on the x-axis and the overpressure on the
y-axis. The crosses
indicate the observed values, reported in \cite{melspa99}, and the
dome height expressed in terms of overpressure.
\par
The values obtained by the best fit of the observed data are:
\be
\begin{array}{lcll}
\displaystyle
\alpha^{best} & = & 3.56 & \hbox{W m$^{-2}$ K$^{-1}$}\\[1ex]
\displaystyle
\mu_w^{best}  & = & 5\times 10^8 & \hbox{Pa s} \\[1ex]
\displaystyle
{\cal B}^{best} & = & 3.46 \\
\end{array}
\label{eq:bestval}
\ee
\figref{fig:fitdati} shows the agreement of the model with
the observed data; the proposed model is able to explain the hysteresis effect
observed in dome growth at Soufriere Hill (Montserrat) by
\cite{melspa99}, and modeled in a different way.
\par
A typical value of the viscosity is $\mu_0=\mu(T_0)=10^7$\,Pa\,s for
Montserrat andesite at $1123\,\hbox{K}$ with 4\% water content
\citep{melspa99}. This value is in good agreement with
\eqref{eq:bestval}; in fact, for ${\cal B}={\cal B}^{best}$ and assuming
$T_w\approx 873\,\hbox{K}$,  we have $\mu_0=\mu_w^{best}\exp{({\cal
B})}\simeq 1.6 \times 10^7$\,Pa\,s. Moreover, for ${\cal B}={\cal
B}^{best}$ and $T_0-T_w\approx 250 \hbox{\ K}$ gives $\beta\simeq 0.014\hbox{\
K$^{-1}$}$  (for example, from data of \cite{hesdin96} for a magma 
with a similar composition and 4\% water
content, we obtain $\beta\simeq 0.016\mbox{\ K$^{-1}$}$). 
Finally, for the heat transfer coefficient we have:
\be
\alpha\approx \frac{k}{\delta_T}\approx 4 \quad\mbox{W m$^{-2}$ K$^{-1}$}
\label{eq:alfa2}
\ee
where $\delta_T$ is the thermal boundary layer of the flow, while using
$k= 2\hbox{\ W m$^{-1}$ K$^{-1}$}$, $\delta_T\approx 50\hbox{\ cm}$
(\cite{bruhup89} used $\delta_T\approx 10\hbox{\ cm}$ for dyke
length of about one kilometer).
\par
Moreover, we verify the basic assumptions of the model:
the assumption of one-dimensional flow is based on the small
diameter/length ratio of the conduit ($R/L \sim 10^{-3}$), and the small
Reynold number is simply verified:
\be
R_e = \frac{\rho R \bar{v}}{\mu_0} \approx \frac{2300 \times 15 \times 0.003}
{10^7} \sim 10^{-5} 
\ee
The viscous heating effects can be neglected because the Nahme number based
on the shear stress is small:
\be
{\cal G}=\frac{\beta(-\frac{dP}{dx})^2 R^4}{4k\mu_w} \approx 1
\label{eq:grunt}
\ee
The assumption of negligible heat conduction along the streamlines is
justified by the high value of the Peclet number (the ratio between
the advective and the conductive heat conduction):
$ P_e = (\rho c_p v R)/k \approx 10^5$.
\par
\begin{figure}
\noindent\includegraphics[angle=-90,width=\figwidth]{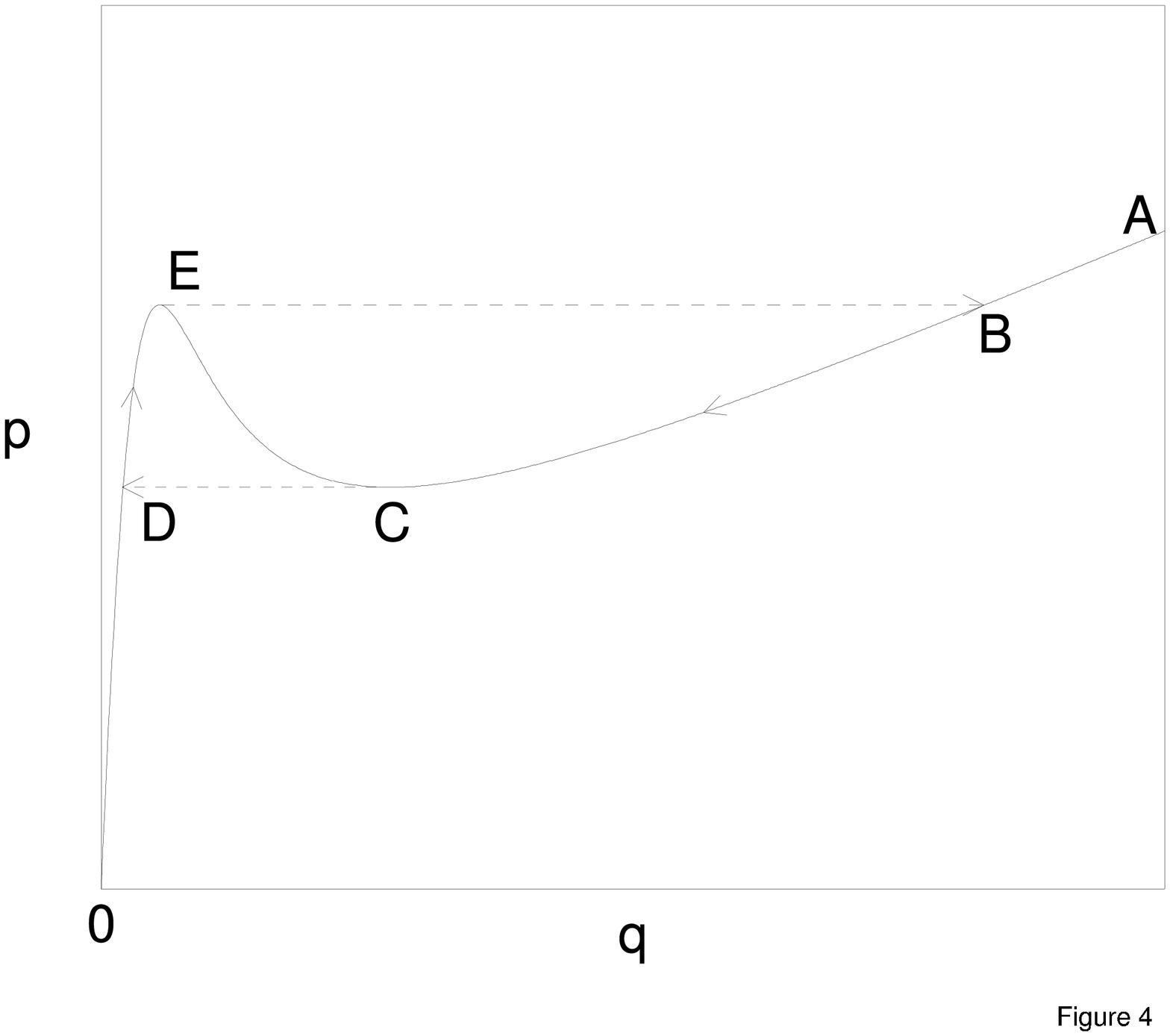}
\caption{Possible pulsating behavior predicted by the model.}
\label{fig:pqfig}
\end{figure}
The existence of the multiple solutions for the steady flow allows the system
to show a pulsating behavior between the different solutions. In the case 
where the initial pressure, is greater than the critical pressure
corresponding to point E in
\callout{\figref{fig:pqfig}}, the system is on the
fast branch of the solution, such as in point A. If the pressure
decreases, the system moves along this branch up to the critical
point C. In C a jump to the slow branch occurs (point D).
If the pressure continues to decrease, the discharge rate
tends to zero. Instead, if the pressure increases, the system moves
along the slow branch up to the other critical point E. In this point
the jump occurs on the fast branch and the system reaches point B.
\par
The overpressure conditions and
pulsating activity, typical of dome eruptions, are evident not only at
Soufri\`ere Hills, but also in Santiaguito (Guatemala), Mount Unzen
(Japan), Lascar (Chile), Galeras (Colombia) and Mount St.\ Helens
(USA) \citep{melspa99}.

\section{Conclusion}
Magma flow in conduits shows the existence of multiple solutions, like
other fluids with strong temperature dependent viscosity. This is a
consequence of the increase of viscosity along the flow due to
cooling. For a given pressure drop along the conduit, one or two stable
regimes (fast and
slow branches) may exist. The transition between the two branches
occurs when critical values are reached, and an hysteresis phenomenon
is possible. These jumps were evident during the dome growth in the
1995-1999 Soufri\`ere Hills (Montserrat) eruption. The pulsating
behaviour of the dome growth was previously modeled by
\cite{melspa99} in terms of the nonlinear effects of crystallization and gas
loss by permeable magma.
\par
In this paper we propose a model to describe the nonlinear
jumps between the two stable solutions as a consequence of the
coupling between the momentum and energy equation induced by the
strong temperature-dependent viscosity of magma.
\par
However, since the crystal content is a consequence of cooling, the
two models, although different, are physically related.
\acknowledgement
This work was supported by the European Commission (Contract
ENV4-CT98-0713), with the contribution of the Gruppo Nazionale per
la Vulcanologia INGV and the Department of the Civil Protection in Italy.
%
% *************** References ***************
%
%\bibliography{references}
%

%
%
\end{article}
\end{document}